\documentclass[lettersize,journal]{IEEEtran}
\usepackage{amsmath,amsfonts}
\usepackage{algorithmic}
\usepackage{algorithm}
\usepackage{array}
\usepackage[caption=false,font=footnotesize,labelfont=rm,textfont=rm]{subfig}
\usepackage{textcomp}
\usepackage{stfloats}
\usepackage{url}
\usepackage{verbatim}
\usepackage{graphicx}
\usepackage{cite}
\newtheorem{lem}{\textit{Lemma}}
\newtheorem{thm}{\textit{Theorem}}

\begin{document}
\title{Sensing-Assisted Beam Tracking with Real-Time Beamwidth Adaptation for THz Communications }
\author{Wuhan Chen, Yuheng Fan, Chuang Yang, \IEEEmembership{Member, IEEE}, and Mugen Peng, \IEEEmembership{Fellow, IEEE}
\thanks{This work is supported in part by National Key Research and Development Program of China under Grant 2021YFB2900200, 2021YFB2900203, in part by the Beijing Natural Science Foundation under grant L223007, in part by the National Natural Science Foundation of China under Grant 61925101, and 62101059 \textit{(Corresponding author: Mugen Peng)}

Wuhan Chen, Yuheng Fan, Chuang Yang, and Mugen Peng are with the State Key Laboratory of Networking and Switching Technology, Beijing University of Posts and Telecommunications, Beijing 100876, China (email: chenwuhan@bupt.cn; vanyh@bupt.edu.cn; chuangyang@bupt.edu.cn; pmg@bupt.edu.cn).

This work has been submitted to the IEEE for possible publication. Copyright may be transferred without notice, after which this version may no longer be accessible.
}}

\markboth{Journal of \LaTeX\ Class Files,~Vol.~18, No.~9, September~2020}%
{How to Use the IEEEtran \LaTeX \ Templates}

\maketitle

\begin{abstract}
Terahertz (THz) communications, with their substantial bandwidth, are essential for meeting the ultra-high data rate demands of emerging high-mobility scenarios such as vehicular-to-everything (V2X) networks. In these contexts, beamwidth adaptation has been explored to address the problem that high-mobility targets frequently move out of the narrow THz beam range. However, existing approaches cannot effectively track targets due to a lack of real-time motion awareness. Consequently, we propose a sensing-assisted beam tracking scheme with real-time beamwidth adaptation. Specifically, the base station (BS) periodically collects prior sensing information to predict the target's motion path by applying a particular motion model. Then, we build a pre-calculated codebook by optimising precoders to align the beamwidth with various predicted target paths, thereby maximising the average achievable data rates within each sensing period. Finally, the BS selects the optimal precoder from the codebook to maintain stable and continuous connectivity. Simulation results show that the proposed scheme significantly improves the rate performance and reduces outage probability compared to existing approaches under various target mobility.
\end{abstract}


\section{Introduction}
\IEEEPARstart{A}{mong} emerging applications for next-generation wireless systems, high mobility scenarios such as unmanned aerial vehicle (UAV) communications \cite{thz_cav} and vehicle-to-everything (V2X) networks \cite{thz_uav} require enhanced network capacity reaching up to Terabits per second (Tbps). To achieve the ultra-high data rate necessary for these applications, recent studies have considered terahertz (THz) communications as one of the key enablers \cite{6g} due to its substantial bandwidth which is typically around 10 gigahertz (GHz). However, obstacles exist in employing THz in high mobility scenarios. THz communications suffer from severe signal attenuation, primarily caused by free space path loss and molecular absorption \cite{a_los}. To compensate for the attenuation, very large antenna arrays are employed to generate directional narrow beams with high beamforming gain. Nonetheless, the narrowness of these beams is likely to induce misalignment, as the target is prone to move out of the effective beamwidth \cite{misalign_1,misalign_2}. Such beam misalignment can cause significant performance degradation which presents major challenges in beam tracking and beamforming. 

Beamwidth adaptation has been explored as an promising solution to counteract misalignment errors by balancing the trade-off between beam coverage and signal strength. Existing studies have employed precoders designed to generate beams with arbitrary predefined width \cite{event_based, multi_res, opt_beam}. Moreover, reference \cite{event_based} has proposed a tracking algorithm triggered by outage events that automatically adapts the beamwidth to enhance reliability. However, given that the achievable data rate is subject to the target's instantaneous velocity \cite{closed_form}, the awareness of the target's real-time mobility is crucial for effective beamforming and tracking in high mobility scenarios. Yet the aforementioned approaches solely operate under a static mobility assumption and cannot adapt beams in response to the real-time target motion.

Furthermore, recent advancements in integrated sensing and communications (ISAC) provide new opportunities in beamforming and tracking, benefiting from the \textit{Coordination Gain} attained partially from sensing-assisted communications \cite{isac,closed_form}. In parallel, in reference \cite{fast_chann}, prior information obtained through conventional channel estimation has been utilised to track the fast-varying THz channels by predicting target motion. Although these studies do not directly address the scenarios in this paper, they inspire us to apply prior sensing assistance obtained by a specific ISAC design for motion prediction. Such an approach can help determine the desirable scale of beam coverage, and thus support effective beam tracking by incorporating real-time motion awareness.

Consequently, in this paper we propose a novel sensing-assisted beam tracking scheme with real-time beamwidth adaptation for THz communications that provides stable and continuous connectivity under various target mobility. Our approach commences by utilising the prior sensing information obtained periodically by the base station (BS) to predict the target's motion path under a specific motion model. Based on the path prediction, the BS selects the optimal beam from a pre-calculated codebook. Furthermore, we optimise precoders with beamwidth adaptation in response to the target's real-time mobility, which aims at maximising the average rate performance while adhering to constraints on minimum instantaneous rate and total transmit power. In particular, we transform the original optimisation problem into a single-variable unconstrained one by employing parameterised precoders and making necessary simplifications. Simulation results demonstrate that our proposed scheme outperforms conventional ones in terms of average achievable rate and outage probability by consciously aligning the beam with the range of target motion. The performance improvement is especially significant at high velocities.

The remainder of this paper is organised as follows. In Section \uppercase\expandafter{\romannumeral2}, we first introduce the THz communication model and then propose the sensing-assisted beam tracking scheme. In Section \uppercase\expandafter{\romannumeral3}, we optimise the mobility-aware precoder. Simulation results are presented in Section \uppercase\expandafter{\romannumeral4}. Finally, conclusions are drawn in Section \uppercase\expandafter{\romannumeral5}.

\textit{Notation}: Boldface letters represent vectors; \(\lvert \cdot \rvert\) denotes the modulus of a complex number; \((\cdot)^T\) and \((\cdot)^H\) denote the transpose and conjugate transpose of a vector respectively; \(\mathcal{CN}(\mu,\Sigma)\) denotes the Gaussian distribution with mean \(\mu\) and covariance \(\Sigma\); \(\mathcal{U}(D)\) denotes the two-dimensional uniform distribution over area \(D\).

\section{System Model}
We consider a typical THz multiple-input multiple-output (MIMO) system in this paper, where the BS employs a uniform linear array (ULA) with \(N_t\) fully digital antennas, and each target employs a single isotropic antenna. Without loss of generality, we assume the targets to be independent of each other due to the sparsity of THz channels \cite{sparsity}. Therefore, we focus solely on the downlink transmission from the BS to a single target for the remainder of this paper.

\subsection{Sensing-Assisted THz Communication Model}
We consider that the BS periodically acquires sensing information such as location, velocity, and acceleration of the target through specific ISAC design, e.g., waveform, signal processing, and hardware design, which has been studied in pertinent works \cite{isac,irs_based,loc_sens,simul}. The sensing period is denoted as \(\tau\). The received signal in the downlink can be written as
\begin{equation}
	y(\boldsymbol{S})=h_0(\boldsymbol{S})\boldsymbol{a}^H(\boldsymbol{S})\boldsymbol{f}x+n
\end{equation}
where \(\boldsymbol{S}=[s_0, v_0, t]\) represents the sensed motion state, with \(s_0\) and \(v_0\) denoting the sensed target location and velocity at the beginning of each period, and \(t \in [0,\tau]\) representing the elapsed time within each period. Here, \(x\) and \(y(\boldsymbol{S})\) are the transmitted and received scalar signal respectively; \(n \sim \mathcal{CN}(0,N_0)\) is the additive Gaussian white noise (AWGN) with \(N_0\) being the noise power spectral density; \(h_0(\boldsymbol{S})\) represents the channel gain; \(\boldsymbol{a}(\boldsymbol{S})\in \mathbb{C}^{N_t\times 1}\) is the array response vector; \(\mathbf{f}\in \mathbb{C}^{N_t\times 1}\) is the precoding vector. The introduction of \(\boldsymbol{S}\) demonstrates the system's real-time responsiveness to target motion, as it encapsulates the target's path prediction during sensing intervals, laying the foundation to optimise the adaptive beamwidth precoder \(\boldsymbol{f}\). Additionally, due to the high scattering attenuation at THz bands, non-line-of-sight (NLoS) channels and multi-path interference are considered negligible in this paper \cite{nlos}.

Without loss of generality, we assume the antenna spacing to be half a wavelength, i.e., \(\Delta=\lambda/2\). The Fraunhofer distance of the array is then defined as
\begin{equation}
	d_{\mathrm{Fraunhofer}}=\frac{2A^2}{\lambda}=\frac{2((N_t-1)\Delta)^2}{\lambda}=\frac{(N_t-1)^2\lambda}{2}
\end{equation}
where \(A\) denotes the aperture length of the ULA. Although hybrid-field conditions have been explored for THz communications \cite{sbarr_based}, this paper focuses on far-field regions. This focus is justified because, according to the parameter settings in Section \uppercase\expandafter{\romannumeral4}, \(d_{\mathrm{Fraunhofer}}\) is smaller than the target's distance from the BS. As a result, planar-wave propagation is assumed, and the array response vector of the ULA is given as 
\begin{equation}
	\boldsymbol{a}(\boldsymbol{S})=[1, e^{-j\pi \mathrm{sin}\phi(\boldsymbol{S})},..., e^{-j(N_t-1)\pi \mathrm{sin}\phi(\boldsymbol{S})}]^T
	\label{arv}
\end{equation}
where \(\phi(\boldsymbol{S})\) is the target's physical direction at time \(t\), with \(\phi(\boldsymbol{S})=0\) if the target is at the broadside direction. Considering free space path loss and molecular absorption, the channel gain is defined as \cite{a_los}
\begin{equation}
	h_0(\boldsymbol{S})=\frac{c}{4\pi d(\boldsymbol{S})f_c}e^{-\frac{K(f_c)}{2}d(\boldsymbol{S})}
	\label{channel_gain}
\end{equation}
where \(c\) is the speed of light, \(d(\boldsymbol{S})\) is the target's distance from the BS at time \(t\), \(f_c\) is the carrier frequency, and \(K(f)\) is the frequency-dependent molecular absorption coefficient. Consequently, the achievable data rate can then be written as 
\begin{equation}
	R(\boldsymbol{S})=B\mathrm{log}_2 \left(1+\frac{P_th_0^2(\boldsymbol{S})}{N_0 B}\lvert \boldsymbol{a}^H(\boldsymbol{S})\boldsymbol{f}\rvert^2 \right)
    \label{r}
\end{equation}
where \(P_t\) is the total transmit power budget, and \(B\) is the bandwidth.

\subsection{Sensing-Assisted Beam Tracking Scheme}

We consider the scenario where a BS with a fixed location tracks a target with two-dimensional motion. We proposed a three-step sensing-assisted beam tracking scheme, detailed as follows.
\begin{enumerate}
    \item \textbf{\textit{Obtaining prior sensing information}}: The BS periodically collects prior sensing information, such as the target's location, velocity, and acceleration.
    \item \textbf{\textit{Predicting target path}}: The BS predicts the target's motion path within a sensing period according to a presumed motion model and the collected prior information.
    \item \textbf{\textit{Selecting optimal beam}}: The BS looks up in a pre-calculated codebook and selects the beam which best covers the target's predicted path and provides the maximum average rate during the sensing period.
\end{enumerate}
The codebook is indexed by angular intervals, with each interval corresponding to a precoder characterised by specific beamwidth and transmit direction. The design of the mobility-aware precoder with beamwidth adaptation will be detailed in Section \uppercase\expandafter{\romannumeral3}. By periodically executing these steps, the BS is expected to effectively track the target and ensure continuous radiation coverage. 

\begin{figure*}[t]
	\center{\includegraphics[width=\linewidth]{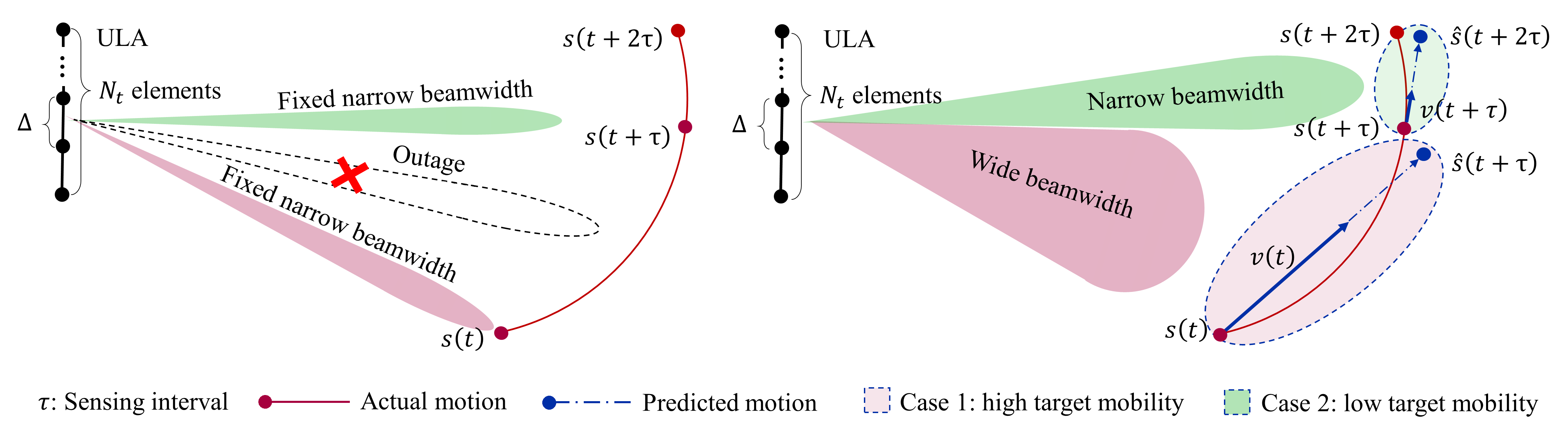}}
	\caption{Models of the THz communication system: the left one uses the conventional scheme, and the right one uses our proposed sensing-assisted beam tracking scheme.}
	\label{sysmod}
\end{figure*}

Fig. \ref{sysmod}, shown at the top of the next page, illustrates the THz communication system where we assume that the target exhibits varying mobility across two adjacent sensing intervals. It also compares our proposed beam tracking scheme with the conventional method. Here, \(\tau\) denotes the sensing interval. The conventional beam tracking scheme solely uses fixed-width beams generated by maximum ratio transmission and redirects the beam towards the target at each \(\tau\) interval, which may result in undesirable outages due to discontinuity in beam coverage.

In Fig. \ref{sysmod}, \(s(t), s(t+\tau),\) and \(s(t+2\tau)\) represent the sensed target locations, which are considered accurate due to the assumption of precise sensing in this work. \(v(t)\) and \(v(t+\tau)\) represent the sensed instantaneous target velocities. \(\hat{s}(t+\tau)\) and \(\hat{s}(t+2\tau)\) represent the predicted target locations under the assumption of \textit{uniform rectilinear motion (URM)}. Note that other motion models, such as uniformly accelerated rectilinear motion, are also compatible with our scheme, with the sole distinction being the kinetic formulation used for motion prediction. However, for simplicity, we exclusively consider URM in this paper. Given the inertia and high mobility of the target, the prediction error is considered negligible, especially when \(\tau\) is small. The BS adjust the beam at every sensing interval \(\tau\). Our proposed scheme dynamically adapts between narrow beams for low target mobility and wide beams for high target mobility, demonstrating an effective balance between coverage and signal strength.




\section{Sensing-Assisted Adaptive Beamwidth Precoder}
In this section, we expect to design the optimal precoder to maximise the achievable data rate while providing extensive coverage of the predicted target path. To this end, we formulate the following optimisation problem
\begin{equation}
	\begin{aligned}
		Q1: &\mathop{\arg\max}\limits_{\boldsymbol{f}} \frac{1}{\tau}\int_{0}^{\tau}R(\boldsymbol{S}|t)\mathrm{d}t \\
		\mathrm{s.t.}\ &C1:\ R(\boldsymbol{S}|t)\geq R_{\mathrm{min}}\\
		&C2:\ \boldsymbol{f}^H\boldsymbol{f}\leq 1
	\end{aligned}
	\label{raw_opt}
\end{equation}
where the objective function is the time average of achievable data rate over a sensing period \(\tau\), \(C1\) is the minimum instantaneous rate constraint introduced to reduce outages caused by beam misalignment, and \(C2\) is the transmit power constraint. Due to its non-convex nature and the fact that the objective function cannot be integrated in closed form, problem (\ref{raw_opt}) is difficult to handle.

However, the proposed sensing-assisted beam tracking scheme allows us to simplify the problem by considering a specific beam shape that features concentration within the predicted range, constant main-lobe beamforming gain, and adaptive beamwidth. More specifically, by constraining the beam shape, we can use limited parameters to control the physical direction and beamwidth, which significantly reduces the dimensionality of the decision variable. To this end, we formulate the precoder as the weighted linear combination of an infinite number of array response vectors within the angular range of predicted target path, which is defined as \cite{event_based}
\begin{equation}
	\boldsymbol{f}(\delta,\omega)=\frac{\beta}{2\delta}\int_{-\delta}^{\delta}\boldsymbol{u}(p,\omega)e^{j\omega p}\mathrm{d}p
	\label{prec}
\end{equation}
where \(\beta\) is the power constraint coefficient, \(\delta\) is introduced to indicate beam coverage, the offset parameter \(\omega\) is introduced to optimise the beam shape, and
\begin{equation}
	\boldsymbol{u}(p,\omega)=[1,e^{-j\pi(p+\theta_m)},...,e^{-j(N_t-1)\pi(p+\theta_m)}]^T
	\label{u_arv}
\end{equation}
is the array response vector at the direction of \(\theta_p=p+\theta_m\). Note that the integral in (\ref{prec}) can lead to unacceptable computational complexity when optimising the precoding vector. Therefore, several simplifications are employed. First, direction expressions are written as the sine of the target's angular direction, transforming the domain from \([-\pi/2,\pi/2]\) to \([-1,1]\). Second, we define
\begin{equation}
	\delta=\frac{-\mathrm{sin}\phi_0+\mathrm{sin}\phi_\tau}{2}
	\label{simp1}
\end{equation}
and 
\begin{equation}
	\theta_m=\frac{\mathrm{sin}\phi_0+\mathrm{sin}\phi_\tau}{2}
	\label{simp2}
\end{equation}
where \(\mathrm{sin}\phi_0\) and \(\mathrm{sin}\phi_\tau\) are the target's directions at time \(0\) and \(\tau\) respectively. Considering (\ref{simp1}) and (\ref{simp2}), we achieve symmetrical bounds of integration in (\ref{prec}), which simplify the calculation and also focus the beam coverage on the target's motion path. In this way, the n-th component of \(\boldsymbol{f}\) can be written as
\begin{equation}
	\begin{aligned}
		f_n&=\frac{\beta}{2\delta}\int_{-\delta}^{\delta}e^{j(\omega-(n-1)\pi)p}e^{-j(n-1)\pi\theta_m}\mathrm{d}p \\
		&=\beta e^{-j(n-1)\pi\theta_m}\mathrm{Sa}(\delta(\omega-(n-1)\pi))
	\end{aligned}
	\label{ele_prec}
\end{equation}
where \(\mathrm{Sa}(x)=\mathrm{sin}(x)/x\) is the sample function. 

Because multi-path interference is considered negligible in this work, it is generally recognised that transmit power is positively related to the achievable data rate under the circumstances. As a result, \(C2\) can be rewritten as
\begin{equation}
	C2':\ \boldsymbol{f}^H\boldsymbol{f}=1
	\label{pw_cnstr}
\end{equation}
For simplicity, given a known \(\delta\), let
\begin{equation}
	g_n(\omega)=\mathrm{Sa}(\delta(\omega-(n-1)\pi))
	\label{sa}
\end{equation}
Then according to (\ref{pw_cnstr}), we have
\begin{equation}
	\beta=\sqrt{\frac{1}{\sum_{n=1}^{N_t}\lvert f_n\rvert^2}}=\sqrt{\frac{1}{\sum_{n=1}^{N_t}g_n^2(\omega)}}
	\label{beta}
\end{equation}
By applying (\ref{beta}) in (\ref{ele_prec}), we actually eliminate the need for \(C2'\) as a separate constraint. Now we are able to reformulate the original problem \(Q1\) into a non-convex single-variable constrained optimisation problem \(Q2\) as follows
\begin{equation}
	\begin{aligned}
		Q2: &\mathop{\arg\max}\limits_{\omega} \frac{1}{\tau}\int_{0}^{\tau}(R(\boldsymbol{S})|t,\omega,\boldsymbol{f}^H\boldsymbol{f}= 1)\mathrm{d}t \\
		\mathrm{s.t.}\ &C1
	\end{aligned}
\end{equation}
The decision variable \(\omega\) helps minimize the variation of beamforming gain within the range of target motion. To address \(C1\) as well as further simplify the problem, we introduce a penalty function
\begin{equation}
	F_p(R(\boldsymbol{S}))=\left\{
	\begin{aligned}
		&-\alpha(R_{\mathrm{min}}-R(\boldsymbol{S})),\ &&\mathrm{if} \ R(\boldsymbol{S})\leq R_{\mathrm{min}}\\
		&0,\ &&\mathrm{else}
	\end{aligned}
	\right.
	\label{penalty}
\end{equation}
and add it to the objective function. The penalty function is non-zero and inversely proportional to the degree of \(C1\) violation when the constraint is violated, and is zero otherwise. The impact of penalty is controlled by the parameter \(\alpha\). By transforming the constrained problem into an unconstrained one with this penalty, we finally obtain
\begin{equation}
	\begin{aligned}
		Q3: &\mathop{\arg\max}\limits_{\omega} \frac{1}{\tau}\int_{0}^{\tau}(R(\boldsymbol{S})+F_p(R(\boldsymbol{S}))|t,\omega,\boldsymbol{f}^H\boldsymbol{f}= 1)\mathrm{d }t
	\end{aligned}
	\label{Q3}
\end{equation}
which can be solved by applying particle swarm optimisation (PSO) \cite{pso}. Although the antiderivative of the integrand in (\ref{Q3}) can not be written in closed form, we opt for numerical integration since the computations are manageable for pre-calculating the codebook.

Next, we need to establish the bounds for PSO. Based on (\ref{arv}), (\ref{ele_prec}), (\ref{sa}), and (\ref{beta}), the beamforming gain can be presented as
\begin{equation}
	\begin{aligned}
		\lvert\boldsymbol{a}^H(\boldsymbol{S})\boldsymbol{f}&(\delta,\omega)\rvert^2=\beta^2 \left(\sum_{m=1}^{N_t}g_m^2(\omega)\right.\\
		&\left.+\sum_{m=n+1}^{N_t}\sum_{n=1}^{N_t-1}2\mathrm{cos}(\Theta_m-\Theta_n)g_m(\omega)g_n(\omega)\right)
	\end{aligned}
	\label{bf_gain}
\end{equation}
where 
\begin{equation}
    \Theta_m=-(m-1)\pi(\mathrm{sin}\phi_m-\phi(\boldsymbol{S}))
\end{equation}
and
\begin{equation}
    \Theta_n=-(n-1)\pi(\mathrm{sin}\phi_m-\phi(\boldsymbol{S}))
\end{equation}
are introduced to simplify the expression.

We present the following \textit{Theorem 1} to demonstrate the axial symmetry of \(\lvert\boldsymbol{a}^H(\boldsymbol{S})\boldsymbol{f}(\delta,\omega)\rvert^2\), which consequently reduces the PSO search space by half.

\begin{thm}
    Given the sensing-assisted information, \(\lvert\boldsymbol{a}^H(\boldsymbol{S})\boldsymbol{f}(\delta,\omega)\rvert\) is symmetric about \(\omega=\frac{N_t-1}{2}\pi\)
\end{thm}

\begin{IEEEproof}
    See Appendix.
\end{IEEEproof}

By (\ref{r}), (\ref{penalty}), and (\ref{Q3}), we can derive from \textit{Theorem 1} that the objective function in problem \(Q3\) is also symmetric about \(\omega=\frac{N_t-1}{2}\pi\). Additionally, it is evident that (\ref{bf_gain}) can be considered zero for \(\omega < 0\), an interval where no optimal case exists. Therefore, we can implement a one-dimensional search using PSO on the interval \([0,\frac{N_t-1}{2}\pi]\) to find the optimal precoder.

\begin{figure}[t]
	\center{\includegraphics[width=0.8\linewidth]{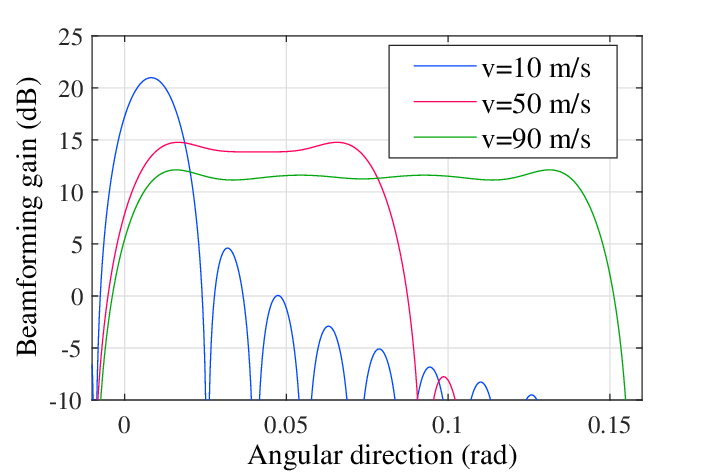}}
	\caption{Optimal beam patterns generated at the velocity of 10 m/s, 50 m/s, and 90 m/s respectively, within a single sensing interval \(\tau\) of 165 ms}
	\label{beam_pattern}
\end{figure}

\section{Simulation Results}
\begin{table}[t]
    \caption{System Parameters for Simulations}
    \centering
    \begin{tabular}{|c|c|}
    \hline
    Number of ULA antennas \(N_t\) & 128\\
    \hline
    Carrier frequency \(f_c\) & 220 GHz\\
    \hline
    Bandwidth \(B\) & 10 GHz\\
    \hline
    Noise power spectrum density \(N_0\) & -174 dBm/Hz\\
    \hline
    Sensing interval \(\tau\) & 165 ms\\
    \hline
    Target motion type & URM\\
    \hline
    Target path & Parallel to ULA\\
    \hline
    Start motion angle& 0 rad\\
    \hline
    End motion angle & 0.3 rad\\
    \hline
    Distance between target path and ULA & 100 m\\
    \hline
    Relative humidity & 50\%\\
    \hline
    Temperature & 14$^{\circ}$C\\
    \hline
    Pressure & 1012.6 hPa\\
    \hline
    \end{tabular}
    \label{simul_param}
\end{table}
In this section, we provide simulation results to evaluate the performance of the proposed sensing-assisted beamforming and tracking scheme in terms of achievable data rate and outage probability. The main simulation parameters are shown in Table \ref{simul_param}. We consider a THz MIMO communication system where the BS is equipped with a 128-antenna ULA and operates at the carrier frequency of 220 GHz with the bandwidth of 10 GHz. The noise power spectrum density \(N_0\) is set to -174 dBm/Hz. To obtain the molecular absorption coefficient \cite{a_los}, we assume a relative humidity of 50\%, a temperature of 14$^{\circ}$C, and a pressure of 1012.6 hPa. We assume the target follows URM. Specifically, the target moves parallel to the ULA at a uniform velocity. The perpendicular distance between the target path and the line on which the ULA is located is set as 100 m. The start angle of target motion is at the broadside direction of ULA which is denoted as \(\phi(\boldsymbol{S}|t=0)= 0\) rad, and the end angle is set to 0.3 rad. This model represents the general case of linear motion, as both the target's angle and distance from the BS vary with the motion. 

Fig. \ref{beam_pattern} presents beam patterns generated by our proposed scheme at various target velocities within a single sensing interval. The sensing interval \(\tau\) is set as 165 ms which will be elucidated later in this section. It can be observed that, aided by prior sensing information, the BS dynamically adjusts beamwidth in response to the target's real-time motion. Narrower beams are utilised for low mobility targets, while wider beams are employed for high mobility targets, though this comes at the cost of reduced main-lobe beamforming gain.
\begin{figure}[!t]
	\centering
	\subfloat[Time average of achievable rate\label{cap_vel}]{
		\includegraphics[width=0.9\linewidth]{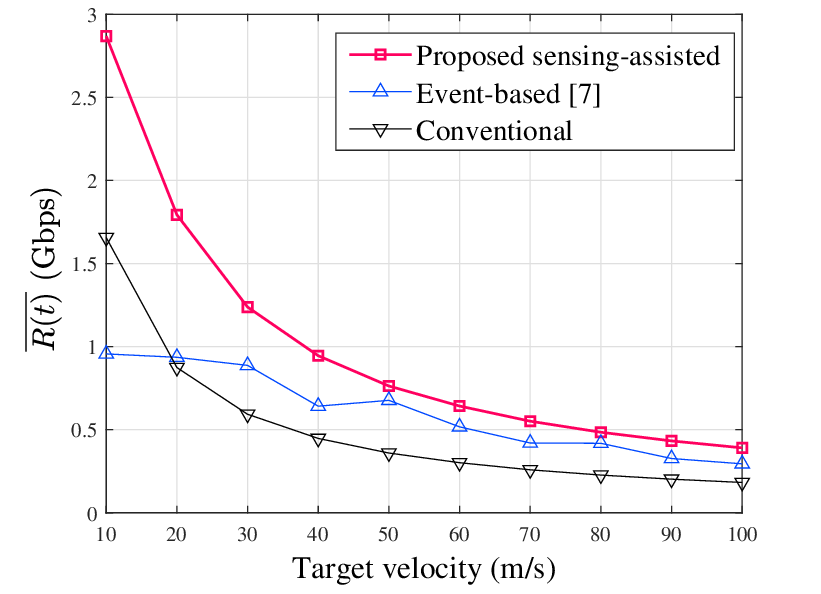}}
	\hfill
	\subfloat[Outage probability\label{out_vel}]{
		\includegraphics[width=0.9\linewidth]{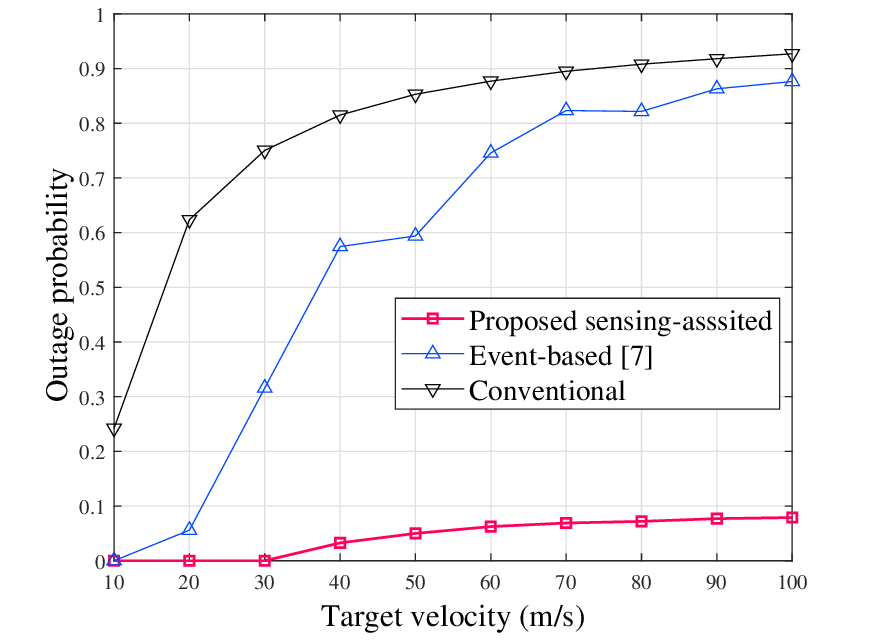}}
        \setlength{\abovecaptionskip}{10pt}
	\caption{Rate and outage performance over the target's motion range of \(\{\phi(\boldsymbol{S})|0<\phi(\boldsymbol{S})<0.3\}\) versus target velocity}
	\label{xaxis_vel}
    \vspace{-11pt}
\end{figure}

\vspace{-11pt}
Then we compare the proposed scheme with the event-based tracking scheme presented in \cite{event_based} and the conventional scheme which we previously introduce in Section \uppercase\expandafter{\romannumeral2}. We set the transmit power to 40 dBm. For reproducing the event-based scheme, we set the time slot duration to 50 ms, the variance of random walk to 25, and the weight parameter to 0.1. We evaluate the performance of different schemes by calculating the time average of achievable date rate and outage probability over the target's motion range \(\{\phi(\boldsymbol{S})|0\leq\phi(\boldsymbol{S})\leq0.3\}\). Note that the beam realignment in \cite{event_based} depends on outages, which occur more frequently with higher target mobility. Consequently, a fair comparison in terms of sensing overhead (or pilot overhead in the conventional scheme) necessitates setting the sensing interval \(\tau\) in our proposed scheme to 3.3 times of the time slot duration in \cite{event_based}. Here, 3.3 represents the average number of elapsed time slots between two adjacent beam realignments, calculated across a range of target velocities from 10 m/s to 100 m/s in 10 m/s increments. Moreover, the interval between beam switches in the conventional scheme is set to the same value as \(\tau\).
\begin{figure}[!t]
	\center{\includegraphics[width=\linewidth]{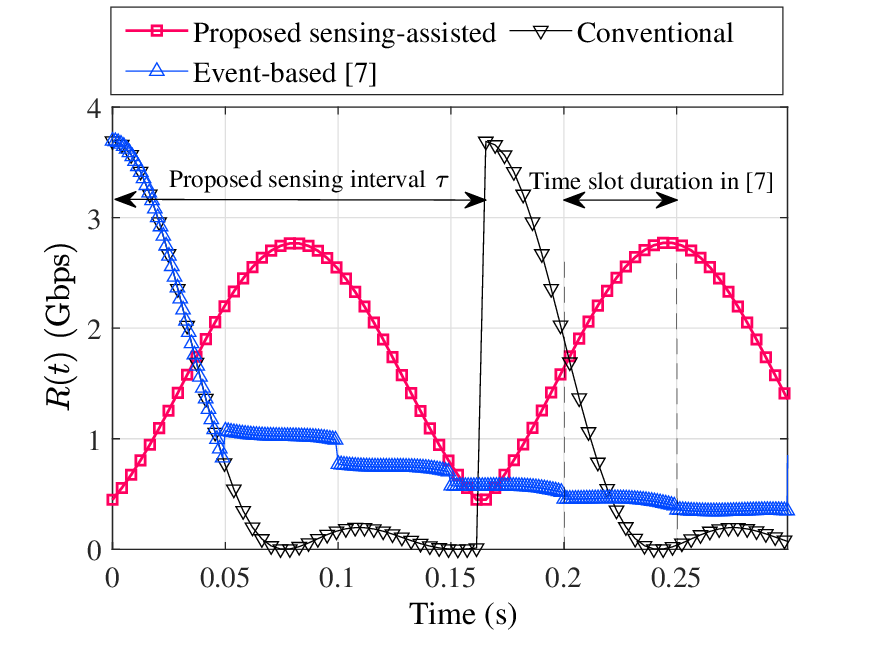}}
        \setlength{\abovecaptionskip}{-12pt}
	\caption{Instantaneous achievable rate with a target velocity of 20 m/s, measured from initialisation to the first beam realignment in the event-based scheme}
	\label{inst_c}
\end{figure}

\begin{figure}[!t]
    \center{\includegraphics[width=0.95\linewidth]{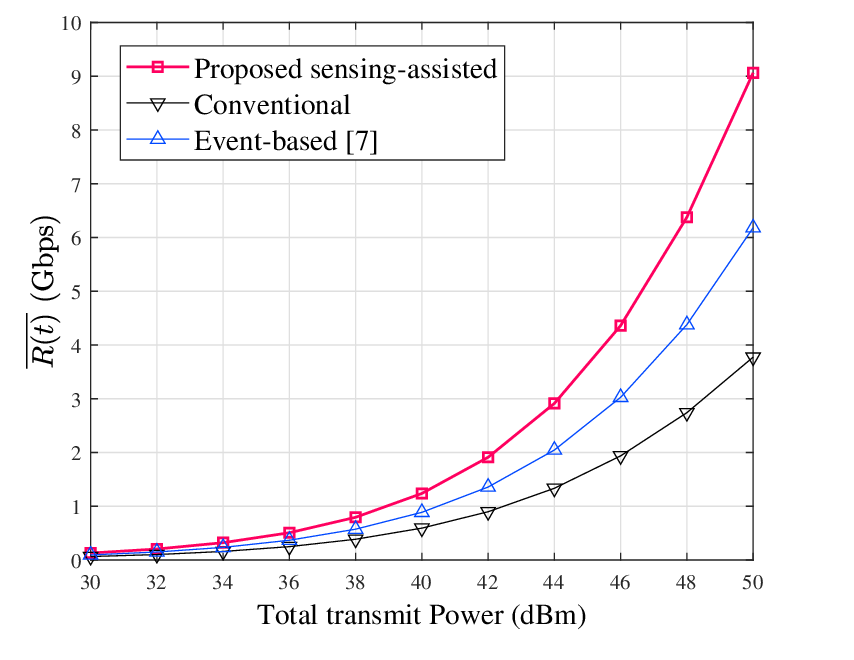}}
    \setlength{\abovecaptionskip}{-3pt}
    \caption{Time average of achievable rate versus total transmit power budget, with a target velocity of 100 m/s}
    \label{cap_Pt}
    \vspace{-9pt}
\end{figure}

Fig. \ref{xaxis_vel} compares the rate and outage performance under various target velocities among different schemes. The range of velocity is set to [10 m/s, 20 m/s, ..., 100 m/s]. We can observe from Fig. \ref{xaxis_vel}\subref{cap_vel} that our proposed sensing-assisted scheme achieves a higher average rate than the other two schemes while there is a downtrend in performance as velocity increases for all schemes. Additionally, Fig. \ref{xaxis_vel}\subref{out_vel} shows that our proposed scheme maintains a lower outage probability, staying below 10\% even at the velocity of 100 m/s. Note that the fluctuation in the event-based scheme's line is due to the non-monotonic decrease in the frequency of beam realignment with increasing velocity.

Fig. \ref{inst_c} presents the instantaneous achievable rate for all three schemes over the time range from the initialisation to the first beam realignment in the event-based scheme. Simulations are run with a target velocity of 20 m/s. We can observe that our sensing-assisted scheme adjusts its beams to fit within the range of target motion, which accounts for its improved rate performance and reduced occurrence of outages. In contrast, the other schemes direct the center of their beams solely towards the target's initial position, leading to almost half of the transmit power being wasted due to the absence of motion awareness. Besides, it illustrates the definition of \(\tau\) and the time slot duration. Further, we can observe that the rate line for our scheme shows approximate axial symmetry within each sensing interval. It is because that, under given conditions, the target's distance from the BS has only trivial effect on achievable rate, which justifies the uniformity of our beam shaping.

Finally, Fig. \ref{cap_Pt} compares the average rate over a range of transmit power for all three schemes at the velocity of 100 m/s, demonstrating the effectiveness of our proposed scheme in high mobility scenarios. 

\section{Conclusions}
Effective beam tracking is crucial in mitigating the performance degradation caused by beam misalignment in THz communications within high-mobility scenarios. In this paper, a sensing-assisted beam tracking scheme with real-time beamwidth adaptation has been proposed, which ensures stable and continuous connectivity for targets with various mobility. Specifically, by periodically collecting the prior sensing information, the BS can predict the path of target motion and then adapt the beamwidth and transmit direction accordingly. To enable such adaptation, we have optimised parameterised precoders that maximise the average achievable rate within each sensing period. Simulation results demonstrate that our scheme significantly outperforms existing methods by achieving improved rate performance and reduced outage occurrence across various mobility conditions, laying the groundwork for future high-mobility applications.

{\appendix[Proof of Theorem 1]
First, we provide \textit{Lemma 1} to prove the axial symmetry of the first summation term on the right-hand side of (\ref{bf_gain}).

\begin{lem}
    \(\sum_{m=1}^{N_t}g^2_m(\omega)\) is symmetric about \(\omega=\frac{N_t-1}{2}\pi\).
\end{lem} 

\begin{IEEEproof}
    Let \(\mu_{k_1,k_2}(x)=\mathrm{Sa}^2(x-k_1)+\mathrm{Sa}^2(x-k_2)\), where \(\mu_{k_1,k_2}(x)\) is an auxiliary function. Then we have
    \begin{equation}
        \begin{aligned}
            &\mu_{k_1,k_2}(x)\\
            =&\mathrm{Sa}^2(-(k_1+k_2-x-k_2))+\mathrm{Sa}^2(-(k_1+k_2-x-k_1))\\
            =&\mathrm{Sa}^2((k_1+k_2-x)-k_2)+\mathrm{Sa}^2((k_1+k_2-x)-k_1)\\
            =&\mu_{k_1,k_2}(k_1+k_2-x)
        \end{aligned}
    \end{equation}
    i.e., \(\mu_{k_1,k_2}(x)\) is symmetric about \(x=\frac{k_1+k_2}{2}\). Similarly, we can prove that \(g^2_{k_1}(\omega)+g^2_{k_2}(\omega)\) is symmetric about \(\omega=\frac{k_1+k_2-2}{2}\pi\). Provided \(N_t\) is even, we have
    \begin{equation}
        \sum_{m=1}^{N_t}g^2_m(\omega)=\sum_{m=1}^{\frac{N_t}{2}}\left(g^2_m(\omega)+g^2_{N_t+1-m}(\omega)\right)
    \end{equation}
    Therefore, we can conclude that \(\sum_{m=1}^{N_t}g^2_m(\omega)\) is symmetric around \(\omega=\frac{N_t-1}{2}\pi\).
\end{IEEEproof}

Subsequently, we provide \textit{Lemma 2} and \textit{Lemma 3} to prove the axial symmetry of the second summation item on the right-hand side of (\ref{bf_gain}). Let the domain of \((m,n)\) be
\begin{equation}
    D=\left\{(m,n)\lvert n\in [1,N_t-1],m\in [n+1,N_t],m,n\in \mathbb{Z}\right\}
\end{equation}
Let
\begin{equation}
    q_{m,n}(\omega)=2\mathrm{cos}(\Theta_m-\Theta_n)g_m(\omega)g_n(\omega)
    \label{q}
\end{equation}
where \(q_{m,n}(\omega)\) is introduced for simplification.

\begin{lem}
    For any \(m,n,m',n'\) that satisfy \(m-n=m'-n'\), there exist a \(\omega_0\) such that \(q_{m,n}(\omega)=q_{m',n'}(\omega+\omega_0)\).
\end{lem}

\begin{IEEEproof}
    Let \(\omega_0=m'-m\) and \(\nu_{m,n}(\omega)=\mathrm{Sa}(\omega-m)\mathrm{Sa}(\omega-n)\), where \(\nu_{m,n}(\omega)\) is an auxiliary function. Then, we have
    \begin{equation}
            \begin{aligned}
                \nu_{m,n}(\omega)&=\mathrm{Sa}(\omega-m)\mathrm{Sa}(\omega-m+(m-n))\\
                &=\mathrm{Sa}(\omega-m'+\omega_0)\mathrm{Sa}(\omega-m'+\omega_0+(m'-n'))\\
                &=\mathrm{Sa}(\omega+\omega_0-m')\mathrm{Sa}(\omega+\omega_0-n')\\
                &=\nu_{m'n'}(\omega+\omega_0)
            \end{aligned}
            \label{lem2_eq1}
    \end{equation}
    Similarly, we can prove that
    \begin{equation}
        g_m(\omega)g_n(\omega)=g_{m'}(\omega+\omega_0)g_{n'}(\omega+\omega_0)
        \label{lem2_eq2}
    \end{equation}
    Additionally, since \(\mathrm{cos}(\Theta_m-\Theta_n)=\mathrm{cos}((m-n)\pi(\mathrm{sin}(\phi_m)-\phi(\boldsymbol{S}))\), we have
    \begin{equation}
        \mathrm{cos}(\Theta_m-\Theta_n)=\mathrm{cos}(\Theta_{m'}-\Theta_{n'})
        \label{lem2_eq3}
    \end{equation}
    Combining (\ref{q}), (\ref{lem2_eq2}), (\ref{lem2_eq3}), we can see that \(q_{m,n}(\omega)=q_{m',n'}(\omega+\omega_0)\).
\end{IEEEproof}

\textit{Lemma 2} implies that for the same value of \(m-n\), the shape of \(q_{m,n}(\omega)\) remains unchanged. Similar to \textit{Lemma 1}, we can prove that \(q_{m,n}(\omega)\) is symmetric about \(\omega=\frac{m+n-2}{2}\pi\), i.e., the position of \(q_{m,n}(\omega)\) is determined by the value of \(m+n\). For random variable \((m,n)\sim \mathcal{U}(D)\), it is trivial to prove that the probability mass function of \(m+n\) is symmetric about \(N_t+1\). Then we prove \textit{Lemma 3} to further present the symmetry of \(\sum_{(m,n)\in D}q_{m,n}(\omega)\). 

\begin{lem}
    Let \(\Delta\) be an integer such that \(\Delta\in[1,N_t-2]\) and \(\Delta\in\mathbb{Z}\). Define the sets \(D_1=\{(m,n)|m+n=N_t+1+\Delta, (m,n)\in D\}\), and \(D_2=\{(m',n')|m'+n'=N_t+1-\Delta, (m',n')\in D\}\). Let \(X_1=m-n\) and \(X_2=m'-n'\) where \((m,n)\sim\mathcal{U}(D_1)\) and \((m',n')\sim\mathcal{U}(D_2)\). Then, for any \(x_0\in\mathbb{Z}\), \(P(X_1=x_0)=P(X_2=x_0)\), 
\end{lem}

\begin{IEEEproof}
    First, we prove the case in which \(\Delta\) is odd. For \((m,n)\sim\mathcal{U}(D_1)\), we have 
    \begin{equation}
        m-n=N_t+1+\Delta-2n
        \label{m-n}
    \end{equation}
    Since \(m>n\) and \(m\leq N_t\), it follows that \(m\in\left[\frac{N_t+1+\Delta}{2}+1,N_t\right], n\in\left[\Delta+1,\frac{N_t+1+\Delta}{2}-1\right]\).
    We denote the upper bound of \(m\) as \(m_{max}\). Thus, for \(n_0\) in the domain of \(n\), the probability \(P(n=n_0)\) can be expressed as
    \begin{equation}
        P(n=n_0)=(m_{max}-n_0)/\sum_{n=\Delta+1}^{\frac{N_t+1+\Delta}{2}-1}(m_{max}-n)
    \end{equation}
    Similarly, for \((m',n')\sim\mathcal{U}(D_2)\), we have 
    \begin{equation}
        m'-n'=N_t+1-\Delta-2n'
        \label{m'-n'}
    \end{equation}
    and \(m'\in\left[\frac{N_t+1-\Delta}{2}+1,N_t-\Delta\right], n'\in\left[1,\frac{N_t+1-\Delta}{2}-1\right]\). Thus, the probability \(P(n'=n'_0)\) can be written as
    \begin{equation}
        P(n'=n'_0)=(m'_{max}-n'_0)/\sum_{n'=1}^{\frac{N_t+1-\Delta}{2}-1}(m'_{max}-n')
    \end{equation}
    Therefore, 
    \begin{equation}
        P(n=n_0)=P(n'=n_0+\Delta)
        \label{P_n}
    \end{equation}
    A similar argument to (\ref{P_n}) holds for the case where \(\Delta\) is even. By (\ref{m-n}), (\ref{m'-n'}), (\ref{P_n}), we can see that the probability distributions of \(X_1\) and \(X_2\) follow
    \begin{equation}
        P(X_1=x_0)=P(X_2=x_0)
    \end{equation}
\end{IEEEproof}

Recall that the position of \(q_{m,n}(\omega)\) is determined \(m+n\) and that the shape of \(q_{m,n}(\omega)\) is determined \(m-n\). \textit{Lemma 3} implies that for every \((m,n)\in D\) that satisfies \(m+n=\rho\), where \(\rho \in \mathbb{Z}\), there exists a corresponding \((m',n')\in D\) that satisfies \(m'+n'=2(N_t+1)-\rho\) and \(m'-n'=m-n\). Consequently, for every \(q_{m,n}(\omega)\), \((m,n)\in D\), there exists a symmetric counterpart \(q_{m',n'}(\omega)\) with \((m',n')\in D\) that is symmetric to \(q_{m,n}(\omega)\) about \(\omega=\frac{N_t+1-2}{2}\pi\) and has the same shape as \(q_{m,n}(\omega)\). Hence, we can conclude that \(\sum_{(m,n)\in D}q_{m,n}(\omega)\) is symmetric about \(\omega=\frac{N_t-1}{2}\pi\).

In summary, the expression on the right-hand side of (\ref{bf_gain}) is symmetric about \(\omega=\frac{N_t-1}{2}\pi\).
}

\bibliographystyle{IEEEtran}
\bibliography{ref}
\vspace{11pt}
\vspace{-33pt}

\begin{IEEEbiography}[{\includegraphics[width=1in,height=1.25in,clip,keepaspectratio]{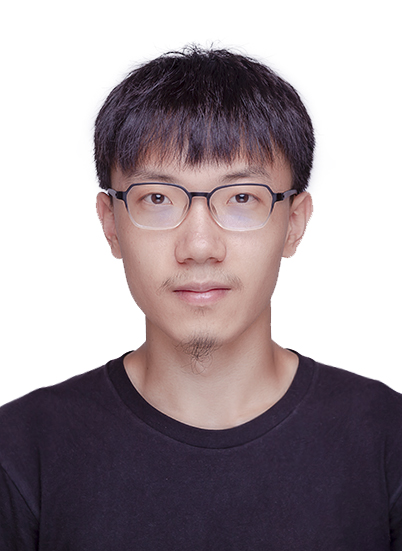}}]{Wuhan Chen}
received the B.S. degree in information and communication engineering from Beijing University of Posts and Telecommunications, Beijing, China in 2024, He is currently pursuing the M.S. degree in communications and signal processing at Imperial College London. His research interests include THz communications, near-field MIMO communications, and ISAC.
\end{IEEEbiography}

\vspace{11pt}
\vspace{-33pt}

\begin{IEEEbiography}[{\includegraphics[width=1in,height=1.25in,clip,keepaspectratio]{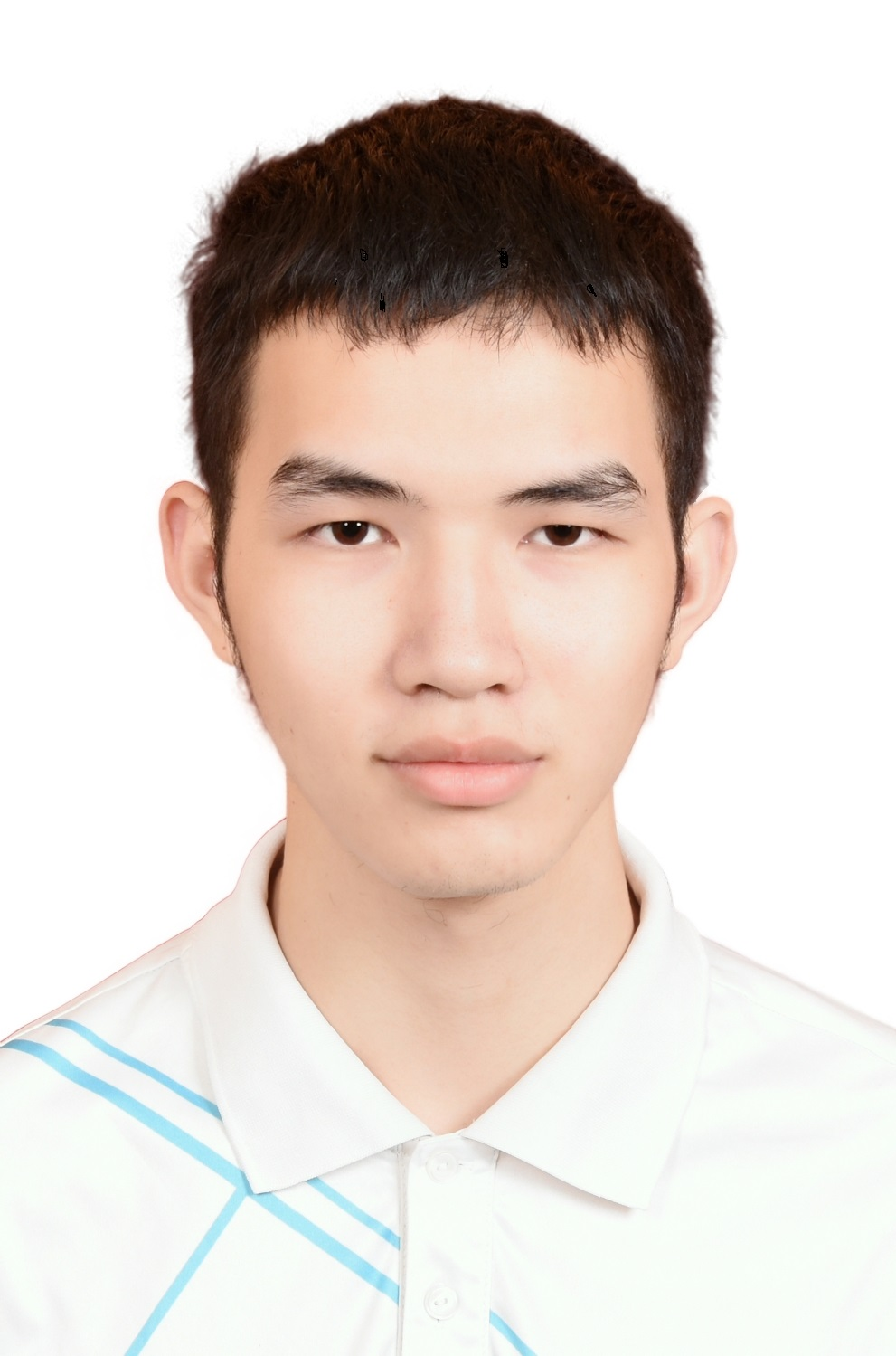}}]{Yuheng Fan}
received the B.S. degree in information and communication engineering from Beijing University of Posts and Telecommunications, Beijing, China in 2023, where he is currently pursuing the Ph.D degree in information and communication engineering with the State Key Laboratory of Networking and Switching Technology. His research is interested in massive MIMO terahertz communication signal processing.
\end{IEEEbiography}

\vspace{11pt}
\vspace{-33pt}

\begin{IEEEbiography}[{\includegraphics[width=1in,height=1.25in,clip,keepaspectratio]{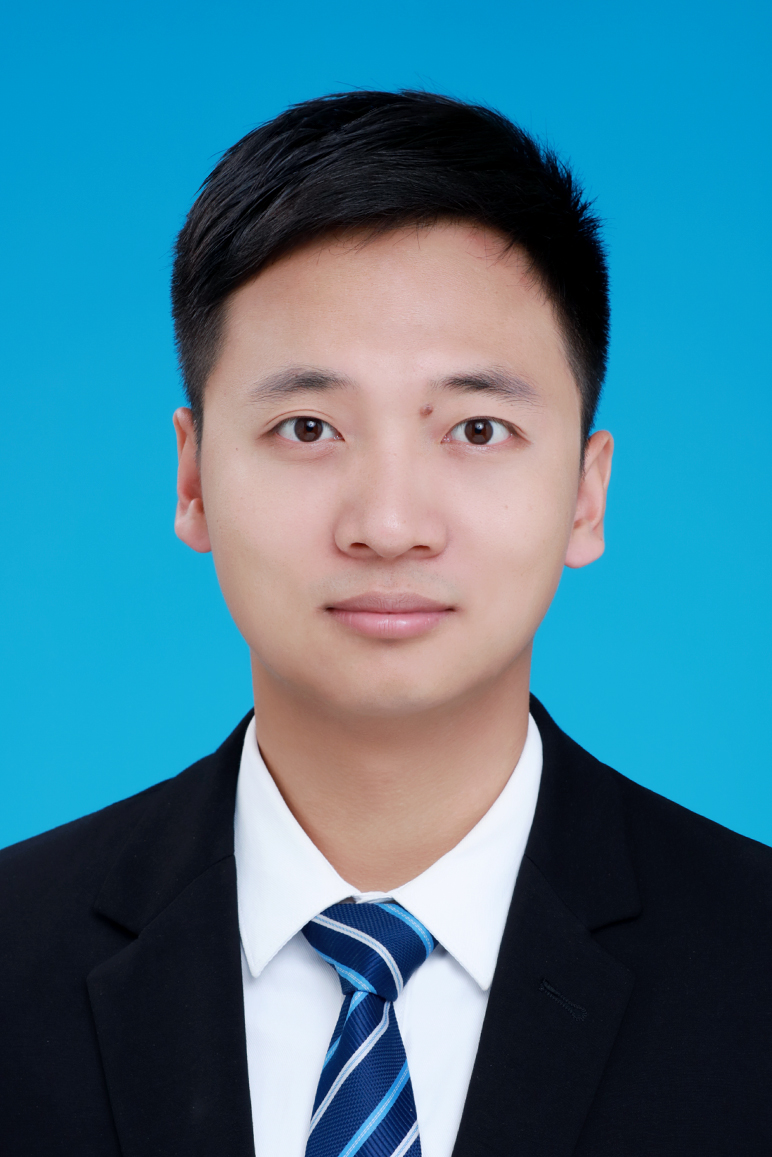}}]{Chuang Yang (Member, IEEE)}
received the B.S. degree in electronic information science and technology from Chongqing University, Chongqing, China, in 2015, and the Ph.D. degree in microelectronics and solid-state electronics from Tianjin University, Tianjin, China, in 2020.
He is currently an Associate Researcher with the State Key Laboratory of Networking and Switching Technology, Beijing University of Posts and Telecommunications, Beijing, China. His research interests include microwave and terahertz measurement, and terahertz integrated sensing and communication.
\end{IEEEbiography}

\vspace{11pt}
\vspace{-33pt}

\begin{IEEEbiography}[{\includegraphics[width=1in,height=1.25in,clip,keepaspectratio]{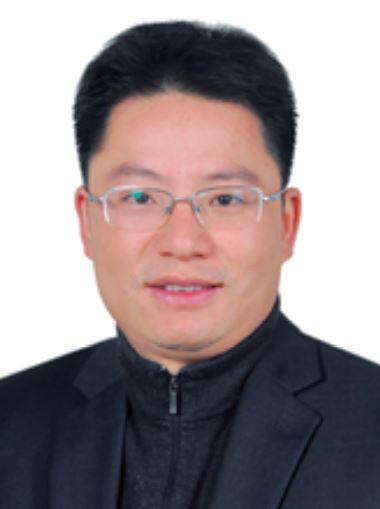}}]{Mugen Peng (Fellow, IEEE)}
received the Ph.D. degree in communication and information systemss from Beijing University of Posts and Telecommunications (BUPT), Beijing, China, in 2005.
Afterward, he joined BUPT, where he has been a Full Professor with the School of Information and Communication Engineering since 2012. In 2014, he was also an Academic Visiting Fellow with Princeton University, Princeton, NJ, USA. He leads a Research Group focusing on wireless transmission and networking technologies with BUPT. He has authored and coauthored over 100 refereed IEEE journal papers and over 300 conference proceeding papers. His main research areas include wireless communication theory, radio signal processing, cooperative communication, self-organization networking, heterogeneous networking, cloud communication, and Internet of Things.
Prof. Peng was a recipient of the 2018 Heinrich Hertz Prize Paper Award; the 2014 IEEE ComSoc AP Outstanding Young Researcher Award; and the Best Paper Award in the ICC 2022, JCN 2016, IEEE WCNC 2015, IEEE GameNets 2014, IEEE CIT 2014, ICCTA 2011, IC-BNMT 2010, and IET CCWMC 2009. He is on the editorial/associate editorial board of the IEEE Communications Magazine, IEEE Access, the IEEE Internet of Things Journal, IET Communications, and China Communications. He is a Fellow of IET.
\end{IEEEbiography}

\vfill

\end{document}